\documentclass[preprintnumbers,nofootinbib]{revtex4}
\usepackage{graphicx}
\usepackage{dcolumn}
\usepackage{bm}
\usepackage{epsfig,amsmath}
\usepackage{amssymb,physics}
\usepackage{subfigure}
\usepackage{float}
\usepackage{ulem}
\usepackage{physics}
\usepackage[usenames,dvipsnames]{color}
\usepackage[pagebackref=false, colorlinks=true]{hyperref}
\definecolor{redish}{rgb}{0.7,0.2,0.0}  
\definecolor{bluish}{rgb}{0.2,0.5,0.8}

\hypersetup{linkcolor=redish,          
                  citecolor=blue,        
                  filecolor=magenta,      
                  urlcolor=bluish}          

\DeclareFontFamily{U}{rsfs}{}         
\DeclareFontShape{U}{rsfs}{m}{n}{<5> rsfs5 <6><7> rsfs7          %
  <8><9><10><10.95><12><14.4><17.28><20.74><24.88> rsfs10}{}     %
\DeclareMathAlphabet{\mathfs}{U}{rsfs}{m}{n}

\def \f{\frac}

\def \o{\omega}

\def \b{\beta}

\def \g{\gamma}

\def \p{\partial}

\def \D{\Delta}

\def \L{\Lambda}

\def \th{\theta}

\def \g{\gamma}

\def \S{\Sigma}

\def \ra{\rightarrow}

\newcommand{\be}{\begin{eqnarray}}
\newcommand{\ee}{\end{eqnarray}}

\begin{document}

\title{Geometric phase in Taub-NUT spacetime}

\author{Chandrachur Chakraborty}
\email{chandrachur.c@manipal.edu} 
\affiliation{Manipal Centre for Natural Sciences, Manipal Academy of Higher Education, Manipal 576104, India}
\author{Banibrata Mukhopadhyay}
\email{bm@iisc.ac.in}
\affiliation{Department of Physics, Indian Institute of Science, Bengaluru 560012, India}

\begin{abstract}
Constructing the Hamiltonian in the $\eta$-representation,
 we explore the geometric phase in the Taub-NUT spacetime, which is spherically symmetric and stationary. The geometric phase around a {\it non-rotating} Taub-NUT spacetime reveals both the gravitational analog of Aharonov-Bohm effect and Pancharatnam-Berry phase, similar to the {\it rotating} Kerr background. On the other hand, only the latter emerges in the spherically symmetric Schwarzschild geometry as well as in the axisymmetric magnetized Schwarzschild geometry. With this result, we argue that the Aharonov-Bohm effect and Pancharatnam-Berry phase both can emerge in the stationary spacetime, whereas only the latter emerges in the static spacetime. 
 We outline plausible measurements of these effects/phases, mostly for primordial black holes.  
\end{abstract}

\maketitle

\section{\label{intro}Introduction}
The Taub-NUT spacetime, described by the two parameters, mass $M$ and NUT (Newman-Unti-Tamburino) charge $n$ \cite{nut}, is geometrically a stationary and spherically symmetric vacuum solution of the Einstein equation \cite{mis}. The violation of the time reflection symmetry ($t \ra -t$) due to the presence of non-zero metric component $g_{t\phi}$ makes the Taub-NUT spacetime stationary.
On the other hand, it possesses three independent spacelike Killing vectors which satisfy the same commutation relations as the Schwarzschild case \cite{mis}. This makes the Taub-NUT spacetime spherically symmetric. 
The NUT charge is also known as the `dual mass' \cite{rs, rs2}, `magnetic mass' \cite{vir}, `gravitomagnetic charge' \cite{cb22}, `gravitomagnetic monopole' \cite{rs2, cbgm}. Bonnor \cite{bon} physically interpreted the NUT charge as `a linear source of pure angular momentum' i.e., `a massless rotating rod' \cite{dow, cb22}. In another paper \cite{abh}, the NUT spcatime was physically interpreted as a static Schwarzschild mass coupled to an externally powered twisting electromagnetic universe.
However, if the NUT charge vanishes, $g_{t\phi}$ becomes zero, and the Taub-NUT spacetime reduces to the Schwarzschild spacetime.

Considering the intriguing nature of NUT parameter, many works have been done in the Taub-NUT spacetime. For example, the Dirac perturbations
of Taub-NUT black holes were studied \cite{abepjc}. The quasinormal modes for the charged Taub-NUT
black holes were also investigated \cite{abks}. Moreover, the chaos for the test particles moving in a Taub-NUT spacetime with a dipolar halo perturbation was studied \cite{lv} and shown that the NUT charge is attenuated in the presence of chaos. The Dirac equation in the Kerr-Taub-NUT spacetime was studied as well \cite{co}, and the Kerr-Taub-NUT spacetime with Maxwell and dilaton fields was explored \cite{acd}.

Lynden-Bell and Nouri-Zonoz were perhaps the first to propose its observational indication in the spectra of supernovae, quasars, and active galactic nuclei \cite{lnbl}. The event horizon telescope (EHT) collaboration also suggested that the super-massive black hole (SMBH) M87* with NUT charge could also be possible \cite{eht5} instead of a Kerr BH. Later, it was shown that a non-zero NUT charge is compatible with the current EHT observations of M87*, that is, the observational constraints on the size and
circularity of the M87* shadow do not exclude the possibility
that this compact object could \cite{gcyl} contain the NUT charge. 
The presence of NUT charge was also indicated in the collapsed object GRO J1655-40 \cite{cbgm}.

Although the Taub-NUT spacetime is spherically symmetric, it is special in this sense that it shows similar effects in many cases, which are generally seen in the axisymmetric spacetimes (e.g., Kerr spacetime). For example, the spin precession of a test gyroscope arisen due to the Lense-Thirring effect can be seen in the Taub-NUT spacetime unlike the Schwarzschild spacetime \cite{cm}. The source of the nonvanishing Lense-Thirring precession leads one to conclude that the Taub-NUT spcetime possesses a `rotational sense' arising due to the non-zero NUT charge. The NUT charge is the one which is responsible to violate the time reflection symmetry. Hence, the spacetime fails to satisfy the hypersurface orthogonality condition, and the neighbouring orbits ``twist'' around each other \cite{wald}. 

In this paper, we are going to explore another interesting effect, i.e., the geometric phase acquired by a spinor in the Taub-NUT spacetime. It is recently shown that the gravitational geometric phase reveals the analog of both the Aharonov-Bohm (AB) effect \cite{ab} and Pancharatnam-Berry (PB) phase \cite{pan, ber} in the axisymmetric Kerr spacetime, whereas the AB effect does not arise in the spherically symmetric Schwarzschild geometry. Here, we show that the AB effect and PB phase both emerge in the Taub-NUT spacetime (similar to the Kerr  spacetime), although the Taub-NUT metric is spherically symmetric. These two widely known classic quantum geometric features are mostly explored in the presence of magnetic fields. 
Assuming the change of environment in an
adiabatic process, Berry showed that in a quantum system, an eigenstate of an instantaneous Hamiltonian returns to its initial state if the Hamiltonian
returns to its initial state, by only acquiring an extra phase factor known as `Berry phase' \cite{ber}. Later, it was shown that the geometric phases appear in a more general ways \cite{sam, aa}. The AB effect, on the other hand, is a quantum mechanical phenomenon in which a charged particle is affected by the electromagnetic potential in spite of the fact that the magnetic field and electric field are zero.

Recently, the similar effects were also shown \cite{gm} to appear in the presence of the gravitomagnetic field which arises due to the motion of `mass' (also called as the `gravitoelectric charge'), i.e., the rotation of the {\it Kerr} black hole. In fact, many similarities (for example, see Introduction of \cite{gm}, \cite{cprd1}) could be found between the general relativity and electromagnetism. Here we mainly focus on the appearance of the geometric phase in the form of AB effect and PB phase for the spinors traversing the gravitational field of Taub-NUT spacetime.

In this paper, we deal with the Dirac Hamiltonian under the gravitational fields in the $\eta$-representation, and investigate the geometric effects of the Dirac particles in the background of Taub-NUT spacetime. As mentioned earlier, an intriguing feature that emerges here is, the Taub-NUT spacetime exhibits the AB effect, in addition to PB phase, similar to the Kerr spacetime. In Sec. \ref{tn}, starting from the basic definition of Taub-NUT metric, we derive the bispinor connectivity using the Schwinger gauge of tetrad. The Dirac Hamiltonian for the Taub-NUT metric in the $\eta-$representation is derived in Sec. \ref{hamil}. The Dirac Hamiltonian in the non-relativistic approximation is deduced in Sec. \ref{nrl}. We derive the Dirac Hamiltonian for the magnetized Schwarzschild spacetime in Sec. \ref{dms} to show that a static and axisymmetric spacetime does not exhibit the AB effect, whereas only the PB phase appears in such a case, as seen in the ordinary Schwarzschild metric. Finally, we conclude and discuss the limitations in Sec. \ref{con}.

\section{\label{tn}Taub-NUT spacetime}
In the geometrized unit\footnote{$G$ is the Newton's gravitational constant and $c$ is the speed of light in vacuum.} ($G=c=1$), the line element of Taub-NUT spacetime is expressed as \cite{manko, vir, bun},
\begin{eqnarray}
 ds^2 &=&- f(r)\left(dt-2n \cos\th d\phi\right)^2+\f{1}{f(r)}dr^2
+ \S (d\th^2+\sin^2\th d\phi^2)
\\
&=& -f(r) dt^2 +\f{1}{f(r)} dr^2 + \S d\th^2 + (\S \sin^2\th-4n^2 f\cos^2\th)d\phi^2+4nf \cos\th dt d\phi,
\label{nut}
\end{eqnarray}
where
\begin{eqnarray}
\D=r^2-2Mr-n^2 \,\,\,\,\, , \,\,\,\,\, \S &=& r^2+n^2 \,\, \,\,\,\,\, ,  \,\,\,\,\, \,\,\,\,\,
 f(r)=\f{r^2-2Mr-n^2}{r^2+n^2}=\f{\D}{\S}.
 \end{eqnarray}
Eq. (\ref{nut}) above indicates that the outer horizon ($r_h^{\rm TN}$) of the Taub-NUT spacetime is located at $r_h^{\rm TN}=M + \sqrt{M^2+n^2}$, whereas the two semi-infinite singularities are located at $\th=0$ and $\th=\pi$ on the symmetry axis \cite{manko}. Note that the detailed discussion on the geometry of the Taub-NUT spacetime could be found in the seminal paper by Misner \cite{mis}.

We choose the Schwinger gauge \cite{gn11, gn14, gm} of tetrad as
\begin{eqnarray}
 e^t_0=\sqrt{-g^{tt}} \,\,\,\, , \,\,\,\,
 e^r_1=\sqrt{\f{\D}{\S}}=\sqrt{f} \,\,\,\, , \,\,\,\,
 e^{\th}_2=\f{1}{\sqrt{\S}} \,\,\,\, , \,\,\,\,
 e^{\phi}_3=\f{1}{\sin\th \sqrt{\D}\sqrt{-g^{tt}}} \,\,\,\, , \,\,\,\,
 e^{\phi}_0=-\f{2n\cot\th}{\S \sin \th \sqrt{-g^{tt}}} ,
\end{eqnarray}
and
\begin{eqnarray}\nonumber
 -e_t^0 &=& e_{0t}=\f{-1}{\sqrt{-g^{tt}}} , \hspace{1cm}
 e_r^1=e_{1r}=\sqrt{\f{\S}{\D}}=\f{1}{\sqrt{f}},\hspace{1cm}
e_{\th}^2= e_{2\th}=\sqrt{\S} ,
\\
e_{\phi}^3 &=&  e_{3\phi}=\sin\th \sqrt{\D}\sqrt{-g^{tt}} , \hspace{1cm}
e_t^3= e_{3t}=\f{2n f \cot\th}{\sqrt{\D}\sqrt{-g^{tt}}} ,
\end{eqnarray}
where 
\begin{eqnarray}
 g^{tt}=-\f{1}{\D}(\S-4n^2 f\cot^2\th)=-\f{1}{f}+\f{4n^2}{\S}\cot^2\th \,\,, \,\,\,\,
g^{t\phi}=\f{2n \cot\th}{\S \sin \th} \,\,, \,\,\,\,
 g^{\phi\phi}=\f{1}{\S \sin^2\th}.
\end{eqnarray}
The Dirac matrices with global indices $\g^{\mu}$ is related to the Dirac matrices with local indices $\g^{a}$ as \cite{gn11, gm} 
\begin{eqnarray}
\g^{\mu} = e^{\mu}_a \g^{a},
\label{gmu}
\end{eqnarray}
where $\mu \equiv (t, r, \th, \phi)$ and $a \equiv (0, 1, 2, 3)$.
Thus, we obtain
\begin{eqnarray}\nonumber
 \g^t &=& e^t_0\g^0=\sqrt{-g^{tt}}\g^0 \,\,,\,\,\,
 \g^r = e^r_1\g^1= \sqrt{f} \g^1 ,
 \\
  \g^{\th} &=& e^{\th}_2\g^2 = \f{1}{\sqrt{\S}} \g^2 \,\,,\,\,\,
   \g^{\phi} =  e^{\phi}_3\g^3 + e^{\phi}_0\g^0 .
\end{eqnarray}
The general expression for the bispinor connectivity $\Phi_{\mu}$ is given by \cite{gn11, gn14, mu05}
\begin{eqnarray}
\Phi_{\mu} &=& \f{1}{4}\o_{bca} S^{bc} e^a_{\mu}
\end{eqnarray}
where $S^{bc} \equiv \f{1}{2} (\g^b \g^c- \g^c \g^b)$ \cite{gn14, gn11} and $\o_{bca}$ are the Ricci rotation coefficients or the spin connections which are calculated in Appendix \ref{rrc}.
Thus, one obtains the components of $\Phi_{\mu}$ for the Taub-NUT spacetime as,
\begin{eqnarray}
\Phi_t &=& \f{1}{2} \left[(\o_{130}e^0_t + \o_{133}e^3_t)\g^1\g^3- (\o_{100}e^0_t + \o_{103}e^3_t)\g^0\g^1 + (\o_{230}e^0_t + \o_{233}e^3_t)\g^2\g^3 - (\o_{200}e^0_t + \o_{203}e^3_t)\g^0\g^2 \right]  , \nonumber
 \label{Phit}
 \\
\Phi_r &=& -\f{1}{2}e^1_r ~ \o_{301}\g^0\g^3 , \nonumber
\label{Phir}
 \\
\Phi_{\th} &=& \f{1}{2}e^2_{\th}\left(\o_{122}\g^1\g^2-\o_{302}\g^0\g^3 \right) , \nonumber
  \\
\Phi_{\phi} &=&  \f{1}{2}e^3_{\phi}\left(\o_{133}\g^1\g^3+\o_{233}\g^2\g^3-\o_{103}\g^0\g^1-\o_{203}\g^0\g^2 \right) .
\label{Phip}
\end{eqnarray}
Note that Eq. (\ref{Phip}) reduces to Eq. (106) of \cite{gn11} for $n \rightarrow 0$.

\section{\label{hamil}Dirac Hamiltonian in the Taub-NUT spacetime}
The general expression for the revised Hamiltonian in the $\eta$ representation $(H_{\eta})$ is given by (see \cite{gn14}, Eq. (89) of \cite{gn11} and Eq. (26) of \cite{gnar})
\begin{eqnarray}
 H_{\eta}=-\f{m}{(-g^{tt})}\g^t+\f{i}{(-g^{tt})}\left[\g^t \g^{\b} \f{\p}{\p x^{\b}}+\g^t \g^{\b}\Phi_{\b}-\f{1}{4}\g^t \g^{\b}\f{\p \ln(g_G g^{tt})}{\p x^{\b}}\right]+\f{i}{4}\f{\p \ln(g_G g^{tt})}{\p t} - i\Phi_t ,
 \label{hgen}
\end{eqnarray}
where $\b \equiv (r, \th, \phi)$, $\eta=(-g_G)^{1/4} (-g^{tt})^{1/4}$ and $g_G=g/g_c$ \cite{gnar}. Note that $g_c$ 
 is the determinant of the matric corresponding to the flat 3-space part, which 
arises when the volume element is written in the curvilinear coordinates \cite{gnar}. For example, $g_c=1$ for Cartesian coordinates, $g_c=r^2$ for
cylindrical coordinates, $g_c=r^4\sin^2\th$ for spherical coordinates etc. For the Taub-NUT spacetime, $g=\det(g_{\mu \nu})=-\S^2 \sin^2\th$ and, hence, $(g_G g^{tt})=(-g^{tt})\S^2/r^4$. Substituting all the necessary expressions in Eq. (\ref{hgen}), the final expression for the Hamiltonian in the $\eta$ representation for the Taub-NUT spacetime is obtained as, 
\begin{eqnarray}\nonumber
H_{\eta} &=& -\f{m\g^0}{\sqrt{(-g^{tt})}}+i\g^0\g^1\f{1}{\sqrt{(-g^{tt})}} \left\{ \sqrt{f}\f{\p}{\p r}+\f{1}{2}(\o_{100}-\o_{122}-\o_{133})\right\}+i\g^0\g^2\f{1}{\sqrt{(-g^{tt})}}\left\{\f{1}{\sqrt{\S}} \f{\p}{\p \th}+\f{1}{2}(\o_{200}-\o_{233})\right\}
\\ \nonumber
&+&i \g^0\g^3 \f{1}{\sqrt{(-g^{tt})g_{\phi\phi}}}\f{\p}{\p \phi}+i\f{2n\cot\th}{(-g^{tt})\S \sin\th}\f{\p}{\p \phi}
-\f{i}{4(-g^{tt})^{3/2}}\left\{\g^0\g^1\sqrt{f}\f{r^4}{\S^2}\f{\p}{\p r}\left(-g^{tt}\f{\S^2}{r^4}\right) + \g^0\g^2\f{1}{\sqrt{\S}}\f{\p}{\p
\th}\left(-g^{tt} \right) \right\}
\\ 
&+& \f{i}{2\sqrt{(-g^{tt})}} (\o_{130}\g^3\g^1+\o_{230}\g^3\g^2).
\label{ham}
\end{eqnarray}
Note, $\g^3\g^2$ and $\g^3\g^1$ can be expressed as $i\g^0\g^1\g^5$ and $-i\g^0\g^2\g^5$ respectively \cite{gm}. Thus, by rearranging Eq. (\ref{ham}) we can write $H_{\eta}$ in a compact form as
\begin{eqnarray}
 H_{\eta}= \f{1}{\sqrt{(-g^{tt})}} \left[-\g^0 m +\g^0\g^j(p_j+iA_j)+i\g^0\g^j\g^5k_j-e^{\phi}_0 p_{\phi} \right],
 \label{gham}
\end{eqnarray}
where
\begin{eqnarray}\nonumber
 A_1 &=& \f{1}{2}(\o_{100}-\o_{122}-\o_{133})-\f{\sqrt{f}}{4(-g^{tt})}\f{r^4}{\S^2}\f{\p}{\p r}\left(-g^{tt}\f{\S^2}{r^4}\right) ,
 \\
 A_2 &=& \f{1}{2}(\o_{200}-\o_{233})-\f{1}{4(-g^{tt})}\f{1}{\sqrt{\S}}\f{\p}{\p
\th}\left(-g^{tt} \right) , \,\,\,\,
 A_3 = 0
  \label{Aj}
\end{eqnarray}
and
\begin{eqnarray}\nonumber
 k_1 &=& \f{i}{2} \o_{230}= -i\f{nf^{1/2}}{2\S g_{\phi\phi}}(2\S-g_{\phi\phi}) ,
 \\
 k_2 &=& -\f{i}{2} \o_{130}= -i\f{n \sin 2\th }{2g_{\phi\phi}\S^{3/2}}(r^3-3Mr^2-3n^2 r+Mn^2) , \,\,\,\, k_3 = 0 .
  \label{kj}
\end{eqnarray}
Here $A_j (\equiv A_1 , A_2 , A_3 )$ is analogous to the magnetic vector potential and $k_j (\equiv k_1 , k_2 , k_3)$ is a `pseudo-vector' potential. We, in general, can call both of them as the `gravitomagnetic potential' \cite{gm} in the Taub-NUT spacetime. The appearance of pseudo-vector potential $k_j$ is an interesting feature in the Taub-NUT geometry, which arises due to the presence of the NUT charge. Note that the gravitomagnetic potentials are the functions of the spacetime coordinates $r$ and $\th$ only.
The intriguing feature that emerges here is, the Hamiltonian (Eq. \ref{gham}) contains a vector and a pseudo-vector
terms, whereas the latter vanishes for the Schwarzschild metric.
Now, writing $H_{\eta}=-i\f{\p}{\p t} \equiv p_t$ and $p_{\mu} \equiv i(-\p_t, \p_r, \p_{\th}, \p_{\phi})$, one obtains 
\begin{eqnarray}
 p_0=e^{\mu}_0p_{\mu}=e^t_0p_t+e^{\phi}_0p_{\phi}=i (-e^t_0\p_t+e^{\phi}_0\p_{\phi}),
 \label{p0}
\end{eqnarray}
and, hence, Eq. (\ref{gham}) reduces to 
\begin{eqnarray}
 \left[p_0 + \g^0 \left\{m - \g^j(p_j+iA_j)-i \g^j\g^5k_j \right\} \right] \Psi=0.
 \label{de}
\end{eqnarray}

\subsection{Special case: $H_{\eta}$ for linear order in $n$}
If we consider the linear order term of $n$ only (neglecting the higher order terms of $n$), Eq. (\ref{ham}) reduces to 
\begin{eqnarray}\nonumber
 H_{\eta}^{\rm lin} &=& -m\sqrt{f_s}\g^0+i\sqrt{f_s}\g^0\left[\g^1\sqrt{f_s}\left(\f{\p}{\p r}+\f{1}{r}\right)+\g^2\f{1}{r}\left(\f{\p}{\p \th}+\f{\cot\th}{2}\right)+\g^3\f{1}{r\sin\th}\f{\p}{\p \phi} \right]+\f{i}{2}\f{\p f_s}{\p r}\g^0\g^1
 \\
&+& i\f{2nf_s\cot\th}{r^2 \sin\th}\f{\p}{\p \phi}-i\f{n\sqrt{f_s}}{r^2}\left[-\left(1-\f{3M}{r} \right)\cot\th \g^3\g^1+ \f{\sqrt{f_s}}{2}(1+2\cot^2\th) \g^3\g^2\right],
 \label{hln}
\end{eqnarray}
where $f_s=(1-2M/r)$. If we explicitly write down the components $A_j$ and $k_j$, equivalently of Eqs. (\ref{Aj}) and (\ref{kj}), for Eq. (\ref{hln}), we obtain
\begin{eqnarray}\nonumber
 A_{1s} = \f{M}{r^2\sqrt{f_s}}+\f{\sqrt{f_s}}{r} ,\hspace{1cm} A_{2s}= \f{\cot\th}{2r},\hspace{1cm} A_{3s} = 0,
\end{eqnarray}
and
\begin{eqnarray}\nonumber
 k_{1s} = -i\f{nf_s^{1/2}}{2r^2}(1+2\cot^2\th) , \hspace{1cm}
 k_{2s} = -i\f{n}{r^2}\left(1-\f{3M}{r} \right) \cot\th ,\hspace{1cm} k_{3s} = 0 .
\end{eqnarray}

\subsection{Special case: $H_{\eta}^{\rm lin}$ for the massless $(M \ra 0)$ Taub-NUT spacetime}
As the Taub-NUT spacetime with $M \ra 0$ is perfectly well-defined \cite{rs2, rs}, one can easily write down the explicit expressions for $H_{\eta}, A_j$ and $k_j$ by setting $M \ra 0$ in Eqs. (\ref{ham}), (\ref{Aj}) and (\ref{kj}) for the massless Taub-NUT spcatime.
However, to visualize the effect of only linear $n$, here we obtain 
\begin{eqnarray}\nonumber
 H_{\eta}^{\rm lin}|_{M \ra 0} &=& -m\g^0+i\g^0\left[\g^1\left(\f{\p}{\p r}+\f{1}{r}\right)+\g^2\f{1}{r}\left(\f{\p}{\p \th}+\f{\cot\th}{2}\right)+\g^3\f{1}{r\sin\th}\f{\p}{\p \phi} \right]
 \\
&+& i\f{2n \cot\th}{r^2 \sin\th}\f{\p}{\p \phi}-i\f{n}{r^2}\left[-\cot\th \g^3\g^1+ \f{1}{2}(1+2\cot^2\th) \g^3\g^2\right]
 \label{hlnm}
\end{eqnarray}
from Eq. (\ref{hln}) by setting $M \ra 0$ and linearizing $n$.
If we explicitly write down the components of $A|_{M \ra 0}$ and $k|_{M \ra 0}$ for Eq. (\ref{hlnm}), we obtain
\begin{eqnarray}\nonumber
 A_{1s}|_{M \ra 0} = \f{1}{r} ,\hspace{1cm} A_{2s}|_{M \ra 0}= \f{\cot\th}{2r},\hspace{1cm} A_{3s}|_{M \ra 0} = 0,
\end{eqnarray}
and
\begin{eqnarray}\nonumber
 k_{1s}|_{M \ra 0} = -i\f{n}{2r^2}(1+2\cot^2\th) , \hspace{1cm}
 k_{2s}|_{M \ra 0} = -i\f{n}{r^2} \cot\th ,\hspace{1cm} k_{3s}|_{M \ra 0} = 0 .
\end{eqnarray}

\subsection{Special case: $H_{\eta}$ for Schwarzschild spacetime}
For $n \rightarrow 0$, Eq. (\ref{ham}) reduces to the $H_{\eta}$ for the Schwarzschild spacetime \cite{gm,gnar}, given by
\begin{eqnarray}
 H_{\eta}^{\rm Sch}=m\sqrt{f_s}\g_0-i\sqrt{f_s}\g_0\left[\g_1\sqrt{f_s}\left(\f{\p}{\p r}+\f{1}{r}\right)+\g_2\f{1}{r}\left(\f{\p}{\p \th}+\f{\cot\th}{2}\right)+\g_3\f{1}{r\sin\th}\f{\p}{\p \phi} \right]-\f{i}{2}\f{\p f_s}{\p r}\g_0\g_1 
 \label{hs}.
\end{eqnarray}
This was already explored in detail earlier \cite{gm}.

\section{Non-Relativistic approximation of Dirac Hamiltonian}
\label{nrl}

Let us now choose the Dirac representation for further calculation: 
\begin{eqnarray}
\gamma^0 =
\begin{pmatrix}
I_2 & 0 \\
0 & -I_2
\end{pmatrix},
\hspace{2cm}
\gamma^i =
\begin{pmatrix}
0 & \sigma^i \\
-\sigma^i & 0
\end{pmatrix},
\hspace{2cm}
\gamma^5 =
\begin{pmatrix}
0 & I_2 \\
I_2 & 0
\end{pmatrix},
\end{eqnarray}
where $I_2$ is a $2 \times 2$ unit matrix and $\sigma^i$ are the Pauli spin matrices. Using Eq. (\ref{de}), we write the Dirac equation as
 \begin{equation}
  \begin{pmatrix}
  -p_0+i  \vec{\sigma} \cdot \vec{k}  -m & \vec{\sigma}\cdot\vec{\Pi} \\
     \vec{\sigma}\cdot\vec{\Pi} & -p_0+i  \vec{\sigma} \cdot \vec{k} +m
  \end{pmatrix} \Psi = 0,
 \end{equation}
where $ \vec{k}=(k_1,k_2,0)$ and $\vec{\Pi}=\vec{p}+i\vec{A}$. It is straightforward to derive the non-relativistic limit of the Hamiltonian following the standard method \cite{sakurai}.
Since $\vec{A}$ and $\vec{k}$ are time independent (as the metric is also time independent), the time dependence of $\Psi$ is given by
\begin{equation}
\Psi = \Psi(\vec{x},t)|_{t=0} e^{-iEt},
\end{equation}
where $E$ is the eigenvalue of the operator $p_0$, with $\Psi$ being the eigenfunction, is given by
\begin{eqnarray}
\Psi = \begin{pmatrix}
\Psi_V \\
\Psi_W
\end{pmatrix}.
\end{eqnarray}
Now, the coupled equations can be written as
\begin{subequations}
\begin{align}
(\vec{\sigma}\cdot \vec{\Pi}) \Psi_W = (E-i\vec{\sigma} \cdot \vec{k} + m) \Psi_V,  \label{eq:subeq1} \\
(\vec{\sigma}\cdot \vec{\Pi}) \Psi_V = (E-i\vec{\sigma} \cdot \vec{k} - m) \Psi_W. \label{eq:subeq2}
\end{align}
\end{subequations}
Substituting $\Psi_V$ from Eq. \eqref{eq:subeq1} in Eq. \eqref{eq:subeq2}, we obtain
\begin{align}
(\vec{\sigma}\cdot \vec{\Pi})\frac{1}{(E-i\vec{\sigma} \cdot \vec{k} + m)}(\vec{\sigma}\cdot \vec{\Pi}) \Psi_W = (E-i\vec{\sigma} \cdot \vec{k} - m) \Psi_W.
\label{coupled_eq_sub}
\end{align} 

Now assuming a small NUT charge and the particles traveling non-relativistically, one can write $ E \sim m$ with $ |\vec{k}|<<m $.
Defining the non-relativistic energy as $E^{NR} = E - m$, we express the above term as
\begin{eqnarray}
\frac{1}{(E-i\vec{\sigma} \cdot \vec{k} + m)} = \frac{1}{2m}\left(\frac{2m}{E^{NR}+2m-i\vec{\sigma}\cdot\vec{k}}\right) 
=\frac{1}{2m}\left(1-\frac{E^{NR}-i\vec{\sigma}\cdot\vec{k}}{2m}+.....\right).
\end{eqnarray}
Keeping only the leading order term, we obtain from Eq. (\ref{coupled_eq_sub})
\begin{eqnarray}
\frac{1}{2m} (\vec{\sigma}\cdot\vec{\Pi})(\vec{\sigma}\cdot\vec{\Pi}) \Psi_W = (E^{NR} - i\vec{\sigma}\cdot\vec{k}) \Psi_W,
\end{eqnarray}
that finally gives
\begin{eqnarray}
\left[\frac{\Pi^2}{2m} + \vec{\sigma} \cdot (i\vec{k}-\frac{i}{2m}\vec{B}_g)\right] \Psi_W = E^{NR} \Psi_W.
\label{hamil_nr}
\end{eqnarray}
Using the Schrodinger equation $H^{NR} \Psi_W=E^{NR} \Psi_W$, we obtain from Eq. (\ref{hamil_nr})
\begin{eqnarray}
\left[\frac{\Pi^2}{2m} + \vec{\sigma} \cdot \vec{B}_g^{\rm NUT}\right] \Psi_W = H^{NR} \Psi_W,
\label{hamil_nr1}
\end{eqnarray}
where $\vec{B}_g=\vec{\nabla}\times\vec{A}$ is the 
effective gravitomagnetic field, and $\vec{B}_g^{\rm NUT}=i(\vec{k}-\frac{1}{2m}\vec{B}_g)$. Thus, it is clear that $\vec{B}_g^{\rm NUT}$  includes a magnetic ﬁeld analogue term $\vec{B}_g$ and a vector
potential analogue term $\vec{k}$.

\subsection{\label{gp}Geometric Phase}
As the case of Dirac particle traveling in a magnetic field, we can write down the interaction between the particle's spin and the background spacetime of the Hamiltonian from Eq. (\ref{hamil_nr1}) as
\begin{equation}
H_{\rm int}=  \vec{\sigma} \cdot \vec{B}_g^{\rm NUT}
\end{equation}
and expand it in a $2 \times 2$ matrix form as
\begin{equation}
H_{\rm int}= |\vec{B}_g^{\rm NUT}|
  \begin{pmatrix}
     \cos\zeta &\sin\zeta \exp(-i\xi) \\
     \sin\zeta \exp(+i\xi) & -\cos\zeta 
  \end{pmatrix},
\end{equation}
where $\zeta$ and $\xi$ are the latitude and azimuthal angles of spherical polar coordinates (see FIG. 1 of \cite{gm}) respectively of the parameter space constructed by $\vec{B}_g^{\rm NUT}$.
To find the geometric phase, let us consider one of the eigenstates of the Hamiltonian following \cite{gm}, i.e.,
\begin{equation}
\ket{\Psi}= 
  \begin{pmatrix}
     -\sin(\frac{\zeta}{2}) \exp(-i\xi)  \\
    \cos(\frac{\zeta}{2})\\
  \end{pmatrix}.
\end{equation}
The phase is defined as \cite{gm}
\begin{align}
 \Phi_P = \int_R i \bra{\Psi} (\hat{\rho}  \frac{\partial}{\partial \rho}  + \hat{\zeta}  \frac{\partial}{\rho \partial \zeta}  + \hat{\xi}  \frac{\partial}{\rho \sin\zeta \partial \xi} ) \ket{\Psi} \cdot d\vec{R},
\end{align}  
where $\rho$ is the radial coordinate, $d\vec{R} = \hat{\rho}d\rho +\hat{\zeta} \rho d\zeta +\hat{\xi} \rho\sin\zeta d\xi$.
The connection is defined as \cite{gm}
\begin{align}
\vec{A}_P= i \bra{\Psi} (\hat{\rho}  \frac{\partial}{\partial \rho}  + \hat{\zeta}  \frac{\partial}{\rho \partial \zeta}  + \hat{\xi}  \frac{\partial}{\rho \sin\zeta \partial \xi} ) \ket{\Psi}.
\end{align} 
Therefore, we obtain \cite{gm}
\begin{equation}
\vec{A}_P = \frac{(1-\cos\zeta)}{2\rho\sin\zeta} \hat{\xi},
\end{equation}
and 
\begin{equation}
\Phi_P =\frac{ \tilde{\xi}}{2} (1-\cos\zeta)=\frac{\Omega}{2},
\label{equation_berry_phase}
\end{equation}
where $\tilde{\xi}$ is the total integrated azimuthal coordinate and
$\Omega$ is the integrated solid angle. Thus, the geometric phase turns out to be of the same known conventional form, e.g., as seen in case of the magnetic field. Note that this is in accordance with the Chern theorem with the Berry curvature $\vec{\nabla}\times \vec{A}_P = \frac{1}{2\rho^3} \vec{\rho} $ \cite{gm} and Chern number $C= \frac{1}{2\pi} \int_S \vec{\nabla}\times \vec{A}_P \cdot d\vec{S} = \frac{\Omega}{4\pi}$ \cite{gm} which is obtained by integrating over the surface $S$. Thus, it is clear that the geometric phase acquired by a quantum system in gravitational background due to the spin interaction with curvature leads to the spin dependent particle dynamics. 

\subsection{Possible measurement}
It is well-known that any quantum effect could be visualized if the underlying length scale $(l)$ is of the order of or less than the corresponding de Broglie wavelength $\lambda = \hbar/p$
of the particle of momentum $p$ with $\hbar$ being the Planck constant. To visualize the geometric effects discussed here, $l$ should typically to be the order of the radius of the gravitating object \cite{gm}. It was recently argued that the speed ($v$) of the electron should be $\lesssim 10^{-11}$ m/s and $\sim 10^{-8}$ m/s in cases of the Earth ($l \sim 6000$ km) and neutron star ($l \sim 10$ km) respectively, which is negligibly small and difficult to detect \cite{gm}. In case of the black hole of mass $M_{\rm bh}$, one can write $l \sim GM_{\rm bh}/c^2$, and obtain
\begin{eqnarray}
 v \sim \f{c}{\sqrt{1+(m_0 M_{\rm bh}/M_P^2)^2}},
\end{eqnarray}
where $m_0$ is the rest mass of the particle (e.g. electron, proton etc.) and $M_P$ is the Planck mass. For the stellar mass black holes, i.e., $M_{\rm bh} \gtrsim 5M_{\odot}$, we obtain $m_0 M_{\rm bh}/M_P^2 >> 1$, and, hence $ v \sim 10^{-10}$ m/s $<< c$ for protons. On the other hand, $v$ could be significantly high (close to $c$) for $M_{\rm bh} \lesssim M_P^2/m_0$. For example, considering $m_0$ as the mass of the electron, $v \sim c$ can be achieved for $M_{\rm bh} \lesssim 10^{15}$ kg, whereas for the proton, the same can be achieved for $M_{\rm bh} \lesssim 10^{12}$ kg. However, $v$ seems to be easily measurable upto $M_{\rm bh} \sim 10^{20}$ kg for what $v\sim 300$ km/sec. The primordial black holes (PBHs), in fact, fall in this mass range. As the PBHs of $M_{\rm bh} > 5 \times 10^{11}$ kg have not been evaporated yet \cite{haw}, the geometric phases can be practically signiﬁcant only for the PBHs of the above-discussed specific mass range, $10^{12}\le M_{\rm bh}/{\rm kg}\le 10^{20}$.

Note that the above discussion is based on the assumption that the PBHs are made of the regular matter/mass $M$ only. Now, if the PBHs contain the NUT charge, the above-mentioned range would be broadened significantly. This is because, it has recently been shown that if a black hole contains the NUT charge $n$ along with the regular matter $M$, the particle production is suppressed by the NUT charge \cite{foo} and, in some cases (see \cite{foo, cb22}), the presence of a small $n$ could even stop the Hawking radiation for a Taub-NUT PBH. If this is true, the PBHs even with $M_P < M_{\rm bh} < 5 \times 10^{11}$ kg could still be found in the present Universe \cite{cb22}. Therefore, the geometric phases in the Taub-NUT spacetime could be signiﬁcant for the PBHs from the Planck mass PBH to $\sim 10^{20}$ kg PBH. It could be useful to mention here that although there is an indication for the existence of NUT charge in nature (see Sec. \ref{intro}), there is no confirmation yet.

Earlier, gravitational Zeeman splitting was proposed by one of the present authors \cite{soumya_universe} for the spinors traversing around a rotating black hole. This would basically arise from the term associated with the AB effect. The present work, hence, argues for the similar effects of splitting of energy levels in the Taub-NUT spacetime, which is also stationary but not static, like the Kerr spacetime. As accreting matter towards a black hole consists predominantly of protons, such an energy splitting is expected in them, governed by the gravitomagnetic potential, not the field. 

\section{\label{dms}Dirac Hamiltonian in the Magnetized Schwarzschild spacetime}
Let us consider an example of the static and axisymmetric spacetime (e.g., magnetized Schwarzschild spacetime) to show that although the PB phase appears in this case, the AB effect does not arise. 
In the geometrized unit, the exact electrovac solution of the Einstein-Maxwell equation for the Schwarzschild BH with a uniform, constant magnetic field\footnote{We should make it clear that $\vec{B}_g$ is the gravitomagnetic field and $\vec{B}$ is the real magnetic field. They are completely different quantities, and there is no relation between $\vec{B}_g$ and $\vec{B}$.} $B$ ($|\vec{B}|$) can be written as  \cite{ernst1976} 
\begin{eqnarray}
 ds^2 &=& \L^2\left[- \left(1-\f{2M}{r} \right)dt^2 + \f{dr^2}{1-\f{2M}{r} }+ r^2 d\th^2 \right]+ \L^{-2}r^2\sin^2 \th d\phi^2,
 \label{sz}
\end{eqnarray}
where
\begin{eqnarray}
 \L=1 + \f{1}{4} B^2r^2 \sin^2\th
  \label{oL}
\end{eqnarray}
  with $r_h=2M $ is the radius of the event horizon. The singularity of this spacetime is located at $r=0$, that is similar to the ordinary Schwarzschild spacetime. Eq. (\ref{sz}), in fact, describes a Schwarzschild black hole which is immersed in an external asymptotically homogeneous and constant magnetic field \cite{gp1978}.

One should note here that the geometry of Eq. (\ref{sz}) is different from the metric considered in Refs. \cite{hay, ovg}, where $g_{tt},~g_{rr},~g_{\th\th}$ do not depend on $\th$. Moreover, the background metric of Refs. \cite{hay, ovg} is independent of $B$ unlike Eq. (\ref{sz}). On the other hand, Ref. \cite{fs} considered a weakly magnetized Schwarzschild black hole and, therefore, the background geometry was still the ordinary Schwarzschild metric, hence is also independent of $B$. Interested readers are referred to Refs. \cite{ernst1976, gp1978} where the geometry of the magnetized Schwarzschild spcatime (Eq. \ref{sz}) are discussed in detail.
  
  The components of the magnetic field in an orthonormal frame are written as \cite{ernst1976}, \cite{gp1978} 
\begin{eqnarray}
 B_r &=& \L^{-2} B \cos\th,
 \\
 B_{\th} &=& -\L^{-2} B\left(1-\f{2M}{r} \right)^{1/2} \sin\th,
\end{eqnarray}
where the angular component vanishes on the event horizon. The magnetized Schwarzschild metric (Eq. \ref{sz}) is not asymptotically flat due to the presence of a strong magnetic field $B$, and it is also not spherically symmetric. Since the magnetic field is assumed to be aligned axially,
the spherical symmetry is broken, and the magnetized Schwarzschild spacetime is considered as axisymmetric \cite{shay, glp}. Moreover, as this metric possesses the time translation symmetry ($t \ra t ~ +$ constant) and time reflection ($t \ra -t$) symmetry \cite{wald}, it is static. 

We choose the Schwinger gauge \cite{gn11, gn14, gm} of tetrad as
\begin{eqnarray}
 e^t_0=\sqrt{-g^{tt}}=\f{1}{\L \sqrt{f_s}} \,\,\,\, , \,\,\,\,
 e^r_1=\f{\sqrt{f_s}}{\L} \,\,\,\, , \,\,\,\,
 e^{\th}_2=\f{1}{r\L} \,\,\,\, , \,\,\,\,
 e^{\phi}_3=\f{\L}{r\sin\th },
\end{eqnarray}
and 
\begin{eqnarray}\nonumber
 -e_t^0 = e_{0t}=-\L \sqrt{f_s} , \hspace{1cm}
 e_r^1=e_{1r}=\f{\L}{\sqrt{f_s}},\hspace{1cm}
e_{\th}^2= e_{2\th}=r\L ,\hspace{1cm}
e_{\phi}^3 =  e_{3\phi}=\f{r\sin\th}{\L}.
\end{eqnarray}
Now, using Eq. (\ref{gmu}) we obtain
\begin{eqnarray}\nonumber
 \g^t = e^t_0\g^0=\sqrt{-g^{tt}}\g^0 \,\,,\,\,\,
 \g^r = e^r_1\g^1= \f{\sqrt{f_s}}{\L} \g^1 \,\,,\,\,\,
  \g^{\th} = e^{\th}_2\g^2 = \f{1}{r\L} \g^2 \,\,,\,\,\,
   \g^{\phi} = e^{\phi}_3\g^3= \f{\L}{r\sin\th } \g^3 ,
\end{eqnarray}
and, thus, we can write
\begin{eqnarray}
\Phi_t &=& -\f{1}{2} e^0_t \left(\o_{100} \g^0\g^1 + \o_{200} \g^0\g^2 \right)  , \nonumber
 \label{Phit}
 \\
\Phi_r &=& \f{1}{2}e^1_r ~ \o_{121}\g^1\g^2 , \nonumber
\label{Phir}
 \\
\Phi_{\th} &=& \f{1}{2}e^2_{\th}~\o_{122}\g^1\g^2  , \nonumber
  \\
\Phi_{\phi} &=&  \f{1}{2}e^3_{\phi}\left(\o_{133}\g^1\g^3+\o_{233}\g^2\g^3 \right) .
\label{PhipB}
\end{eqnarray}
The expressions for $\o_{bca}$ for the magnetized Schawarzschild spacetime are available in Appendix \ref{rrc2}. It would be useful to note here that $\Phi_r$ does not vanish for the magnetized Schwarzschild geometry unlike the ordinary Schwarzschild metric. However, Eq. (\ref{PhipB}) reduces to Eq. (106) of \cite{gn11} for $B \rightarrow 0$.

For the magnetized Schawarzschild metric,  $g=\det(g_{\mu \nu})=-\L^4 r^4 \sin^2\th$ and $g_c=r^4 \sin^2\th$ and, hence, $(g_G g^{tt})=\L^2/f_s$. Now, using Eq. (\ref{hgen}) we obtain $H_{\eta}$ for the magnetized Schwarzschild (mSch) spacetime as
\begin{eqnarray}
 H_{\eta}^{\rm mSch} &=& \L m\sqrt{f_s}\g_0-i\sqrt{f_s}\g_0\left[\g_1\sqrt{f_s}\left(\f{\p}{\p r}+\f{1}{r}\right)+\g_2\f{1}{r}\left(\f{\p}{\p \th}+\f{\cot\th}{2}\right)+\g_3\f{\L^2}{r\sin\th}\f{\p}{\p \phi} \right]-\f{i}{2}\f{\p f_s}{\p r}\g_0\g_1 ,
 \label{hms}
 \\
& \equiv & \f{1}{\sqrt{(-g^{tt})}} \left[-\g^0 m +\g^0\g^j(p_j+iA_j^{\rm mSch}) \right],\nonumber
\end{eqnarray}
where 
\begin{eqnarray}
 A_1^{\rm mSch}= \f{1}{\L}\left[\f{\sqrt{f_s}}{r}+\f{1}{2\sqrt{f_s}}\f{\p f_s}{\p r} \right], \hspace{.5cm} A_2^{\rm mSch}=\f{\cot\th}{2 \L r}, \hspace{.5cm}  {\rm and} \hspace{.5cm} A_3^{\rm mSch}=0.
 \label{ajms}
\end{eqnarray}
Eq. (\ref{ajms}) reveals that although the gravitomagnetic potential $A_j^{\rm mSch}$ is the function of the real magnetic field $B$, the first and second terms within the square bracket and the last term of the Hamiltonian (Eq. \ref{hms}) do not contain any term related of $B$. They look exactly the same as the terms to $H_{\eta}^{\rm Sch}$. Therefore, Eq. (\ref{hms}) can also be written in the form of
\begin{eqnarray}
 H_{\eta}^{\rm mSch}=H_{\eta}^{\rm Sch}+(\L -1) \sqrt{f_s}\g_0 \left[m- i \g_3 \f{(\L + 1)}{r\sin\th}\f{\p}{\p \phi} \right].
  \label{hms1}
\end{eqnarray}
For $B \ra 0$ (i.e.,  $\L \ra 1$), Eq. (\ref{hms}) and Eq. (\ref{hms1}) reduce to  $H_{\eta}^{\rm Sch}$ \cite{gm,gnar} which is also given in Eq. (\ref{hs}).

\subsection{Special case: $H_{\eta}^{\rm mSch}|_{M \ra 0}$ for the Melvin's magnetic Universe}
It is well-known that for $M \ra 0$, Eq. (\ref{sz}) reduces to the metric which represents the Melvin's magnetic Universe (mU) \cite{ernst1976}. This spacetime was extensively studied earlier \cite{melvin, thorne}. However, Eq. (\ref{hms}) reduces to 
\begin{eqnarray}
 H_{\eta}^{\rm mU} \equiv H_{\eta}^{\rm mSch}|_{M \ra 0} = \L m \g_0-i\g_0\left[\g_1\left(\f{\p}{\p r}+\f{1}{r}\right)+\g_2\f{1}{r}\left(\f{\p}{\p \th}+\f{\cot\th}{2}\right)+\g_3\f{\L^2}{r\sin\th}\f{\p}{\p \phi} \right]
 \label{hms0}
\end{eqnarray}
for $M \ra 0$, which can be considered as the Hamiltonian in the $\eta$-representation ($H_{\eta}^{\rm mU}$) for the Melvin's magnetic Universe. Using Eq. (\ref{hms0}), the three components of the gravitomagnetic potential ($A_j^{\rm mU}$) can be written as
\begin{eqnarray}
 A_1^{\rm mU}= \f{1}{\L r}, \hspace{.5cm} A_2^{\rm mU}=\f{\cot\th}{2 \L r}, \hspace{.5cm}  {\rm and} \hspace{.5cm} A_3^{\rm mU}=0.
 \label{ajms0}
\end{eqnarray}
The above non-zero values of $A_j^{\rm mU}$ (Eq. \ref{ajms0}), in fact, indicate that the spinors traversing in the Melvin's magnetic Universe can acquire a non-zero gravitomagnetic potential arisen from a real magnetic field $B$. Thereby, the effective gravitoamgnetic field $\vec{B}_g=\vec{\nabla} \times \vec{A}$ would be a function of a real magnetic field, and the (magnetic-)energy density $(\sim B^2)$ of the real magnetic field contributes to the gravitational mass of the spacetime \cite{bon60} in this case. That is why, we obtain a non-zero gravitomagnetic field and potential as functions of $B$. 

\section{\label{con}Conclusion and Discussion}
In this paper, we have shown that the geometric phases in the form of AB effect and PB phase are appeared in the case of spinors traversing in the gravitational ﬁeld of the Taub-NUT spacetime, and only the latter emerges for the magnetized Schwarzschild spacetime. The main geometric difference between two spacetimes is, the first one is spherically symmetric but stationary \cite{mis}, whereas the second one is axisymmetric but static \cite{shay}. This peculiarity makes them special. The similarity between two spacetimes is that both of them are asymptotically non-flat. Due to this reason, they are not astrophysically so feasible until now. There are also other issues specifically related to the Taub-NUT spacetime, the details of which and their remedies have been vividly discussed by one of us earlier \cite{cb22, gcyl}. 

The main point which we emphasize here is that the gravitational analog of AB effect and PB phase both can arise even in a spherically symmetric, but not static, spacetime if it is stationary. On the other hand, the only PB phase can emerge even in an axisymmetric spacetime, if it is static. Therefore, it is realized that the geometric phases are not specifically related to the symmetry of space, i.e., spherical symmetry and/or axisymmetry, rather, they are related to the symmetry of time, i.e., time translation symmetry and time reflection symmetry. A stationary but nonstatic metric must have $dt dx^{\mu}$ cross terms in any coordinate system which uses the Killing parameter as one of the coordinates \cite{wald} (see Sec. 6.1 of \cite{wald} for detail). Therefore, although it possesses the time translation symmetry ($t \ra t ~ +$ constant), it violates the time reflection ($t \ra -t$) symmetry. This also indicates the violation of the Frobenius's theorem of hypersurface orthogonality \cite{wald}. For the static metric, the cross terms $dt dx^{\mu}$ do not arise, and hence it possesses both the time reflection and time translation symmetry. Thus, it satisfies Frobenius's theorem of hypersurface orthogonality \cite{wald}. Physically, the fields which are time translation invariant can fail to be time reflection invariant, if any rotational motion (e.g., Kerr spacetime) \cite{wald} or, at least, a sort of `rotational sense' (e.g., Taub-NUT spacetime) \cite{cm} is involved. Finally, we conclude that the presence of the cross term $dt dx^{\mu}$ in the metric (as discussed above), for which a spacetime fails to satisfy the hypersurface orthogonality condition, is solely responsible for emerging the AB effect.
\\

{\bf Acknowledgements :} The authors thank Kaushik Ghosh of Vivekananda College and Parthasarathi Majumdar of Indian Association for the Cultivation of Science for discussions. One of the authors (BM) also thanks SERB, India, for partial support by a project with Ref. No. CRG/2022/003460. CC dedicates this paper to his mother Shree Snigdha Chakraborty who was his constant inspiration in all fields of life, and unfortunately passed away on September 11, 2023.

\appendix
 \begin{appendix}
 \section{\label{rrc}Ricci rotation coefficients for the Taub-NUT spacetime}
The Ricci rotation coefficients or the spin connection for the Taub-NUT spacetime are calculated as:
\begin{eqnarray}
 \o_{122}&=&-\f{r f^{1/2}}{\S}=-\o_{212}
 \label{of}
 \\
  \o_{133}&=& -\f{f^{1/2}}{g_{\phi\phi} \S^2}\left[ r\S^2 \sin^2\th -4n^2 w \cos^2\th \right]=-\o_{313}
  \\
  \o_{130}&=& \f{n \sin 2\th }{g_{\phi\phi}\S^{3/2}}(r^3-3Mr^2-3n^2 r+Mn^2)=\o_{103}=\o_{301}=-\o_{013}=-\o_{310}=-\o_{031}
  \\
  \o_{100}&=& \f{1}{g_{\phi\phi}\S^2\sqrt{\S \D} }\left[ w \S^2 \sin^2\th -4rn^2\D^2 \cos^2\th \right]=-\o_{010}
\\
\o_{233}&=&-\f{ (\S+4n^2f)\sin 2\th}{2g_{\phi\phi}\S^{1/2}} =-\o_{323}
\\
\o_{230}&=& \f{nf^{1/2}}{\S g_{\phi\phi}}(g_{\phi\phi}-2\S) =\o_{203}=\o_{302}=-\o_{320}=-\o_{023}=-\o_{032}
\\
\o_{200}&=& \f{4n^2f \cot\th}{g_{\phi\phi}\S^{1/2}} =-\o_{020}
\label{ol}
\end{eqnarray}
where $w=Mr^2+2n^2 r-Mn^2$.

\section{\label{rrc2}Ricci rotation coefficients for the magnetized Schwarzschild spacetime}
\begin{eqnarray}
\o_{121}&=& \f{B^2 r\sin 2\th}{4 \L^2} =-\o_{211}=\o_{200}=-\o_{020}
\label{of2}
\\
 \o_{122}&=& -\f{\sqrt{f_s}}{r \L^2} \left(1+\f{3}{4}B^2 r^2\sin^2\th\right) =-\o_{212}
 \\
 \o_{133}&=& -\f{\sqrt{f_s}}{r \L^2} \left(1-\f{1}{4}B^2 r^2\sin^2\th \right)=-\o_{313}
  \\
  \o_{100}&=&  \f{1}{r^2 \L^2 \sqrt{f_s}}\left[M-\f{1}{4}B^2 r^2\sin^2\th ~(3M-2r) \right]=-\o_{010}
\\
\o_{233}&=& - \f{\cot\th}{r\L^2}\left(1-\f{1}{4}B^2 r^2\sin^2\th \right)=-\o_{323}.
\label{ol2}
\end{eqnarray}

Eqs. (\ref{of}-\ref{ol}) and Eqs. (\ref{of2}-\ref{ol2}) reduce to
\begin{eqnarray}
 \o_{122}&=& -\f{\sqrt{f_s}}{r}  =-\o_{212}
 \\
 \o_{133}&=& -\f{\sqrt{f_s}}{r} =-\o_{313}
  \\
  \o_{100}&=&  \f{M}{r^2\sqrt{f_s}}=-\o_{010}
\\
\o_{233}&=& - \f{\cot\th}{r}=-\o_{323}
\end{eqnarray}
for $n \rightarrow 0$ and $B \rightarrow 0$ respectively,
which are considered as the spin connections for the Schwarzschild spacetime (see Eq. 2.2.8 of \cite{muller}).
\end{appendix}

\end{document}